\documentclass[preprint2]{aastex}
\usepackage{graphicx}
\usepackage{ifthen}

\def\ltsima{$\; \buildrel < \over \sim \;$}
\def\simlt{\lower.5ex\hbox{\ltsima}}
\def\gtsima{$\; \buildrel > \over \sim \;$}
\def\simgt{\lower.5ex\hbox{\gtsima}}
\def\kms{{\rm\,km\,s^{-1}}}

\def\kpc{{\rm\,kpc}}

\def\msun{{\rm\,M_\odot}}

\def\CompactFigs{1}
\def\UseFigs{1}

\def\s{\ifmmode \widetilde \else \~\fi}
\def\={\overline}

\def\spose#1{\hbox to 0pt{#1\hss}}

\def\etal{{\it et al.\ }}
\def\cf{{\it cf.\ }}

\def\lta{\mathrel{\spose{\lower 3pt\hbox{$\mathchar"218$}}
     \raise 2.0pt\hbox{$\mathchar"13C$}}}
\def\gta{\mathrel{\spose{\lower 3pt\hbox{$\mathchar"218$}}
     \raise 2.0pt\hbox{$\mathchar"13E$}}}
\def\Dt{\spose{\raise 1.5ex\hbox{\hskip3pt$\mathchar"201$}}}    
\def\dt{\spose{\raise 1.0ex\hbox{\hskip2pt$\mathchar"201$}}}    

\def\dotsfill{\leaders\hbox to 1em{\hss.\hss}\hfill}

\def\Gyr{{\rm\,Gyr}}

\shorttitle{Ibata et al.}
\shortauthors{Halo substructure in the SDSS}

\begin{document}

\title{Galactic Halo substructure in the Sloan Digital Sky
Survey:\\
the ancient tidal stream from the Sagittarius dwarf galaxy}


\author{
Rodrigo Ibata\altaffilmark{1}, 
Michael Irwin\altaffilmark{2},
Geraint F. Lewis\altaffilmark{3},
Andrea Stolte\altaffilmark{1}}

\altaffiltext{1}{Max-Plank Institut f\"ur Astronomie, 
K\"onigstuhl 17, D--69117 Heidelberg, Germany}
\altaffiltext{2}{
Institute of Astronomy, Madingley Road, Cambridge, CB3 0HA, U.K.}
\altaffiltext{3}{
Anglo-Australian Observatory, PO Box 296, Epping, NSW 1710, Australia}

\begin{abstract}
Two  studies  have  recently  reported  the  discovery  of  pronounced  Halo
substructure  in the  Sloan Digital  Sky Survey  (SDSS)  commissioning data.
Here we  show that this Halo substructure  is almost in its  entirety due to
the expected tidal  stream torn off the Sagittarius  dwarf galaxy during the
course of its many close encounters with the Milky Way.  This interpretation
makes strong  predictions on  the kinematics and  distances of  these stream
stars.  Comparison of the structure in old horizontal branch stars, detected
by  the SDSS  team, with  the carbon  star structure  discovered in  our own
survey, indicates  that this halo stream  is of comparable age  to the Milky
Way. It  would appear that  the Milky Way  and the Sagittarius  dwarf galaxy
have been a  strongly interacting system for most  of their existence.  Once
complete, the SDSS will provide a unique dataset with which to constrain the
dynamical evolution of  the Sagittarius dwarf galaxy, it  will also strongly
constrain the mass distribution of the outer Milky Way.
\end{abstract}




%

\section{Introduction}

In the standard hierarchical picture  of galaxy formation, galaxies like the
Milky Way  grew by repeated mergers  with other galaxies,  and in particular
with many  smaller galaxy ``fragments''.   In the outer regions  of galactic
halos, disruption timescales can be long,  and we can hope to observe galaxy
remnants as  well as  to catch  galaxies in the  process of  being accreted.
That this accretion  is still an on-going process is  clear, as evidenced by
the  many studies  of tidal  features in  the dwarf  satellite  galaxies and
globular clusters.

However, what is of prime interest  is to compare the observed structures to
the results of galaxy formation  theories. These now make strong predictions
about the  structure and substructure  of galaxy halos. Indeed  the standard
cold  dark  matter  models  appear  to be  inconsistent  with  observational
constraints,   as   they    over-predict   halo   clumpiness   (see,   e.g.,
\citealt{moore}).

In an  attempt to  constrain such theories,  several groups  have undertaken
large  scale surveys  to uncover  substructure in  the halo.  These include,
among others,  the ``spaghetti''  survey \citep{morrison00}, the  APM carbon
star  survey  \citep{totten98,  totten00,  ibata00},  and  the  SDSS  survey
\citep{yanny, ivezic}.

The APM carbon star survey, an almost all high-latitude sky survey, revealed
the existence  of a gigantic band  of these intermediate  age ($\sim 6\Gyr$)
stars around  the sky (\citealt{ibata00};  hereafter paper~1). 
\footnote{Extant   optical   and  NIR   photometry   and  detailed   optical
spectroscopy of  the APM  colour-selected cool Halo  carbon stars  show that
these stars  are identical in  nature to those  found in AGB  populations in
other Local  Group galaxies, particuarly  the Magellanic Clouds and  many of
the  dwarf   spheroidal  Galactic  satellites   \citep{totten98,  totten00}.
Furthermore the central  part of Sgr has a known  AGB carbon star population
\citep{whitelock96}   that    also   has   NIR    and   optical   properties
indistinguishable from  the APM  Halo sample.  Normal  AGB carbon  stars are
carbon-enriched during  dredge-ups driven  by helium burning  thermal pulses
during the post  main sequence phase of 1-2 solar mass  stars.  As such they
are thought to be  $> 1-2$~Gyrs old but less than 10~Gyrs old (see e.g. the
reviews by Da Costa 1997 and  also Azzopardi \& Lequeux 1992).}
Approximately half of the ``Halo''  carbon stars detected in this survey are
located in  this ten  degree wide  band, which is  centered on  the expected
orbit  of  the  Sagittarius  dwarf  galaxy in  a  spherical  Halo  potential
(paper~1). Sagittarius, the closest dwarf  satellite galaxy to the Milky Way
\citep{ibata94}, is clearly in the  process of being tidally torn apart.  As
shown in paper~1, those carbon stars trace a gigantic tidal stream of debris
disrupted  off the  Sagittarius dwarf  galaxy which  has now  wrapped itself
around the Halo.   Interestingly, this stream is observed  to follow a great
circle on the sky, which means that it is probably a coplanar rosette-shaped
structure  that we  observe from  within (since  it should  trace relatively
closely  the  orbit of  the  center  of mass  of  the  dwarf galaxy).   Most
importantly,  however, the  fact that  the stream  is confined  to  a plane,
implies  that it  does  not  suffer significant  precession,  which is  only
possible if the Galactic halo potential is not significantly oblate.

Two  SDSS teams  have recently  presented  results from  their first  year's
commissioning  data   \citep{yanny,  ivezic}.   Due   to  initial  technical
constraints,  the drift-scanning  camera  took data  only  on the  celestial
equator during the commissioning period.  Their data consists of two wedges:
a NGC  sample $87^\circ$ long  and a SGC  sample $60^\circ$ long,  with each
covering  a declination  range  of only  $-1.25^\circ  \simlt \delta  \simlt
1.25^\circ$.  Despite  the fact  that their data  provide essentially  a two
dimensional slice through the  Galaxy, they find astonishing sub-structure in
the distribution of A-stars.   Sharply-defined concentric rings of stars are
detected.   The  number  of  RR-lyrae  variables also  cuts  off  sharply  at
$50\kpc$.   The nature of  these structures  is important,  as they  are the
dominant structure seen in old  Halo stars, comprising approximately half of
the total number of Halo stars seen in the survey.  In this contribution, we
show that these A-star  Halo sub-structures are morphologically very similar
to  the expected  structure of  the tidal  debris of  the  Sagittarius dwarf
galaxy, are  at similar distances  and are highly  likely to be part  of the
Sagittarius stream.

At first sight it may seem  surprising that such a limited survey can detect
Halo  sub-structure.   However,  since  most long-lived  Galactic  satellite
systems  appear to be  on close  to polar  orbits the  chances of  a stellar
debris stream  intersecting the current SDSS  zero-declination survey strips
is  $\approx$50\%.  The  Sagittarius stellar  stream is  currently  the only
known large  relatively intact debris system  and is within  $13^\circ$ of a
polar orbit.

\section{The expected debris of the Sagittarius dwarf galaxy}

In  paper~1, we  undertook a  large suite  of numerical  simulations  of the
interaction between the Sagittarius dwarf galaxy and the Milky Way, in order
to determine  how the  APM carbon  star sample constrains  the shape  of the
Galactic  halo.  Figure~1 shows  a typical  high resolution  simulation with
$10^5$ particles, at  the end of the integration,  at $T=12\Gyr$ (details of
these simulations  are given in  paper~1).  For this  particular simulation,
the initial structure  of the dwarf galaxy is a fluffy  King model with mass
$10^8\msun$, half-mass radius of $r_{1/2}=0.5\kpc$ and initial concentration
$c=0.5$.  The  Milky Way model (termed model  H1 in paper~1) has  a bulge, a
thin  disk,  a  thick  disk  and an  interstellar  medium,  with  parameters
identical  to the  \citet{dehnen}  model~2;  the Galactic  halo  has a  flat
rotation curve, with  a constant density core of  radius $r_0=3\kpc$, and an
oblate spheroidal  density structure, with density  flattening of $q_m=0.9$.
The total  mass of the  Galaxy model is  such that the circular  velocity at
$50\kpc$ is $v_c=220\kms$.

After  accounting for  a  $\sim 50$\%  ``background''  from other  accretion
events,  such  simulations  can  give  a reasonable  representation  of  the
observed  carbon  star  distribution,   both  in  spatial  position  and  in
kinematics (\cf paper~1).  The Aitoff projection of this simulation onto the
sky (Figure~1), shows  that the majority of the particles  in this model are
clustered  close  to a  very  obvious  great  circle stream.   However,  the
potential  is not  quite spherical,  so orbits  suffer some  precession. The
material first stripped from the dwarf  galaxy has had longest to drift away
from  its progenitor,  and so  also  experiences the  largest difference  in
precession.   The particles  not on  the great  stream are  those  that were
stripped from the dwarf galaxy at earlier times; indeed, the precession away
from the great circle stream  effectively provides a chronometer of the time
since particles were disrupted from their parent galaxy.

To compare the simulation of Figure~1 to the SDSS commissioning run data, we
take a  $2.5^\circ$ equatorial cut  though the simulation.  As  discussed by
\citet{yanny} and  \citet{ivezic}, there  may be some  mixture of  blue star
populations in their  dataset, however, for simplicity, we  assume that they
are   all   blue   horizontal   branch   stars   with   absolute   magnitude
$M_{g^\star}=1.0$.  The resulting distribution  of these equatorial stars as
a function of right ascension  and apparent $g^\star$ magnitude is displayed
in  Figure~2.  The  resemblance  of  this distribution  with  that found  by
\citet{yanny} (their Figures~3 and  11) is striking.  This simulation (which
preceded the  SDSS result)  predicts the azimuthal  rings seen by  the Sloan
Survey. Figure~3 is identical to  Figure~2, but the simulation it is derived
from was undertaken in a more flattened halo, with $q_m=0.5$. The sharp ring
features  seen  in Figure~2  and  in the  SDSS  dataset  are smeared-out  in
Figure~3.  A  spherical Halo ($q_m=1.0$)  gives rise to a  slower precession
rate (there is still some precession due to the modelled disk component), so
a   smaller   fraction   of   particles   wander  off   the   great   circle
stream. Unfortunately, we  cannot be quantitative in our  comparisons to the
models without a published list of SDSS A-stars.

The color-coding  of Figures~2  and 3 indicates  the time elapsed  since the
particles left the tidal radius  of their progenitor. In Figure~2, the tight
clumps  seen  near  2~hours  and  14~hours are  due  to  recently  disrupted
material, which  can be  composed of young  as well  as old stars.   This is
consistent with  the detection of a  clump of carbon stars  in these regions
(paper~1).  The  particles that were  disrupted long ago, displayed  here in
yellow and  red, have had  many Gyrs to  precess away from the  great circle
stream.  The  only remaining  visible tracers of  this material will  be old
stars,  such as  RRLyrae variables.   This model  suggests that  the ancient
disruption and star-formation history of the Sagittarius dwarf may be traced
and  analysed with  the population  of RRLyrae  variables first  detected by
SDSS.  In the main  body of  the Sagittarius  dwarf galaxy,  blue horizontal
branch stars make up a much  higher fraction of the total stellar population
in its globular  clusters than in its field  stars \citep{layden}.  The fact
that  the  older   parts  of  its  tidal  stream   display  apparently  high
concentrations  of   RRLyrae  stars  suggests  a  common   origin  with  the
Sagittarius  dwarf's  globular  clusters   (though  this  may  just  reflect
continual star-formation in the dwarf galaxy).

Our simulations make strong predictions.   First, the radial velocity of the
A-stars should vary smoothly as  a function of right ascension, as displayed
on the top panel of Figure~1.  The radial velocity dispersion of these stars
in a  small range in  right ascension should  be small also,  $\sim 20\kms$,
except in regions  where multiple streams cross each  other, at $\alpha \sim
2$~hours  and $\alpha  \sim  14$~hours.  Furthermore,  the  majority of  the
A-stars should have  similar radial velocities to those  of the carbon stars
in the 2~hour and 14~hour regions of paper~1. On the equatorial plane of the
sky (where the extant SDSS data  lie) at $\alpha \sim 2$~hours, the velocity
distribution  should be  bimodal,  with two  clumps  at Heliocentric  radial
velocity  of $v \sim  +200\kms$ and  $v \sim  -150\kms$.  Further  along the
equatorial plane at $\alpha \sim 14$~hours the mean Heliocentric velocity of
the stream  stars should  be $\sim 50\kms$,  with significant  dispersion of
$\sigma  \sim  60\kms$.   Secondly,   the  distances  of  the  A-stars,  and
RR-lyraes, should also closely follow  the range of distances found from the
C-star studies.   In particular, an upper  cutoff of $\approx$50  kpc in the
RR-lyrae distances is exactly what is seen in the C-star distribution.

\section{Conclusions}

We have shown  that numerical simulations of the  dynamical evolution of the
Sagittarius  dwarf  galaxy,  previously  reported in  paper~1,  predict  the
presence of sharp shell-like features in the Galactic halo.  This shell-like
structure is due  to the effect of orbital precession  acting on stars which
were  disrupted  from  the  Sagittarius  dwarf  galaxy  many  billion  years
ago. Making a  cut through our simulated dataset  along the equatorial plane
of the sky, we find a distribution  very similar to that found in A-stars in
the SDSS data \citep{yanny, ivezic}.

The  A-stars furthest  away from  the dominant  stream detected  in paper~1,
belong  to  those populations  that  were  first  disrupted from  the  dwarf
galaxy. By comparing  the relative densities of young to  old stars in these
tidal structures, it may be  possible to reconstruct the dynamical evolution
of the Sagittarius dwarf galaxy stream in great detail \citep{johnston}.

The SDSS data that will soon  become available along other scan-lines on the
sky offer the  very useful opportunity to distinguish  between the different
dynamical models of the Sagittarius  dwarf galaxy. Such data will also allow
us to  distinguish competing  models for the  mass distribution of  the dark
halo. For instance, a comparison  of the particle distributions of Figures~2
and 3,  shows that the enhanced  stream precession in the  flattened halo of
Figure~3 gives rise  to a substantial density between  $270^\circ < \alpha <
0^\circ$, which is not seen in  Figure~2. Data in that right ascension range
will provide a means to further constrain the shape of the Galactic halo.

With  the almost  all  high-latitude sky  coverage  of the  APM carbon  star
survey, only  two Halo streams were  found (the other probably  being due to
the Large Magellanic Cloud). Given  that SDSS has also intersected the tidal
stream from the  Sagittarius dwarf galaxy, this stream appears  to be by far
the dominant  contributor to  Halo substructure at  Galactocentric distances
less  than  $\sim 50\kpc$.  It  seems  unlikely,  therefore, that  any  more
accreted systems of  such magnitude will be found by  SDSS or other surveys.
However,  there  is  hope  to  discover the  much  smaller  accreted  galaxy
fragments, if they exist.

The simulation result of Figure~1, has a cautionary tale to tell about using
starcounts  to determine  the shape  of  the ``Halo''.   That a  substantial
fraction  of the  outer  Galactic halo  stars  appear to  be  clumped on  an
inclined band  on the  sky clearly  does not mean  that the  Halo is  a thin
inclined flattened band! Such structures just tell a complicated story about
the process of stellar accretion into the Halo.  Since stars only contribute
a miniscule amount to  the total halo mass, the only way  to reveal the dark
matter distribution  is through dynamical  analyses.  For this  purpose, the
dynamically cold streams discussed in this paper are ideal.

\onecolumn

\begin{figure}
\ifthenelse{\UseFigs=1}{
\ifthenelse{\CompactFigs=0}
{\includegraphics[width=15cm]{Sgr_SDSS.fig01.ps}}
{\includegraphics[width=15cm]{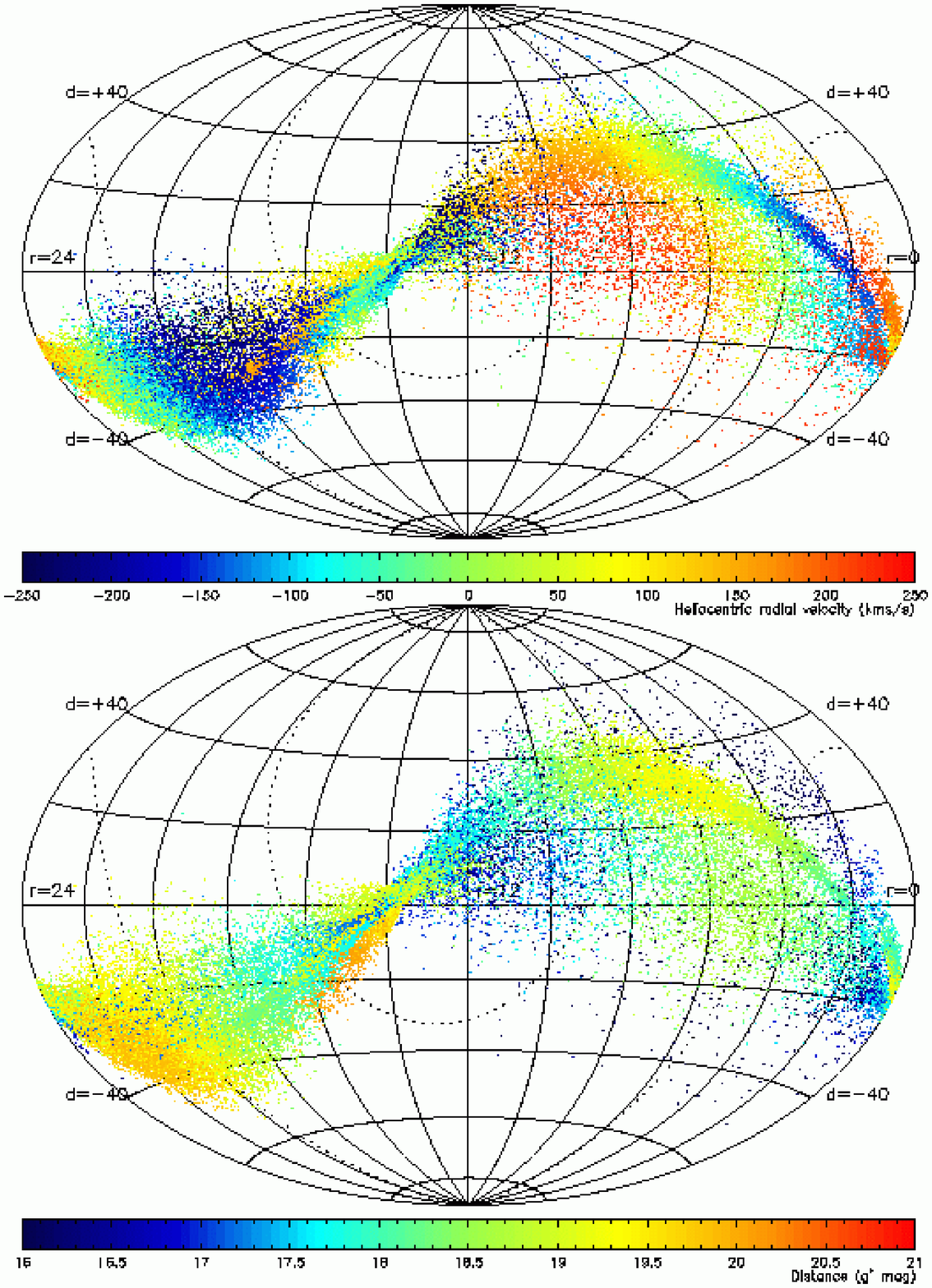}}}{}
\figcaption{The  end-state of the  simulation (at  $T=12\Gyr$) of  the dwarf
galaxy  model D1 (originally  containing $10^5$  particles) in  the Galactic
potential model  H1 with $v_c=220\kms$,  and $q_m=0.9$ is shown.   The upper
panel displays color-coded Heliocentric  radial velocities, while the bottom
panel shows  the stellar distances,  color-coded as a function  of $g^\star$
apparent magnitude.  Though  the distribution of the majority  of the debris
follows approximately  a Great  Circle on the  sky, a significant  amount of
material  is also precessed  away from  the Great  Circle stream.   The SDSS
commissioning run  dataset is confined to $\pm1.25^\circ$  of the equatorial
plane in two stripes at $23.5^h < RA < 3.5^h$ and $9.7^h < RA < 15.5^h$. The
model predicts  a high density  of stream stars  in the SDSS dataset  at two
regions at  right ascension of $\sim  2$~hours and $\sim  14$~hours. This is
where the A-colored star clumps are found in the SDSS dataset.}
\end{figure}

\begin{figure}
\ifthenelse{\UseFigs=1}{
\includegraphics[width=12cm]{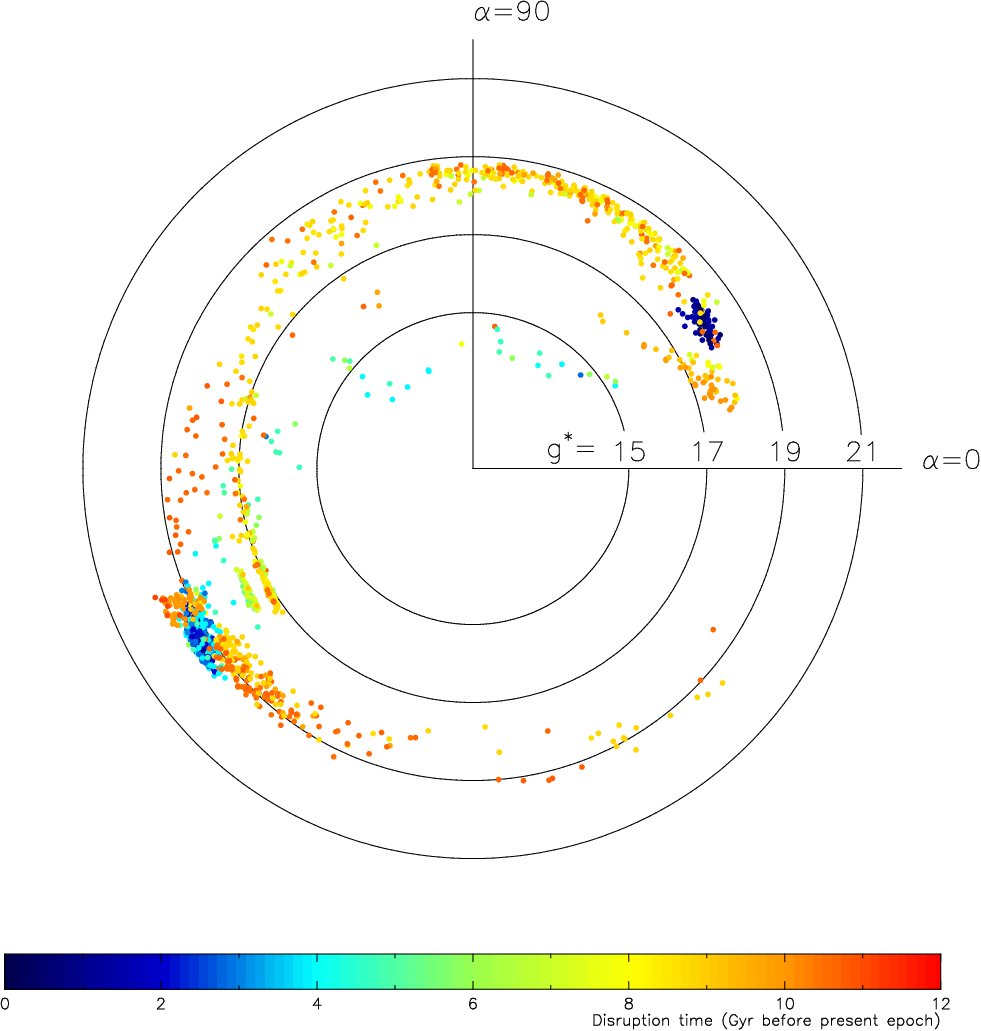}}
\figcaption{The  diagram  shows  a  $2.5$  degree  wide  slice  through  the
(celestial) equatorial plane of  the simulation displayed in Figure~1, which
was  evolved in a  Galactic potential  with an  almost spherical  halo (mass
flattening $q_m=0.9$). The color of  the particles marks the time since they
were ejected  beyond the  tidal limit of  the dwarf galaxy  progenitor.  The
similarity  of  this  distribution  to  the  data  of  \citet{yanny}  (their
Figures~3  and 11)  is striking.   In  this model,  the azimuthal  ring-like
features are  produced by  material that was  disrupted off  the Sagittarius
dwarf galaxy at early times, and which has had a long time to precess in the
Galactic potential.}
\end{figure}

\begin{figure}
\ifthenelse{\UseFigs=1}{
\includegraphics[width=12cm]{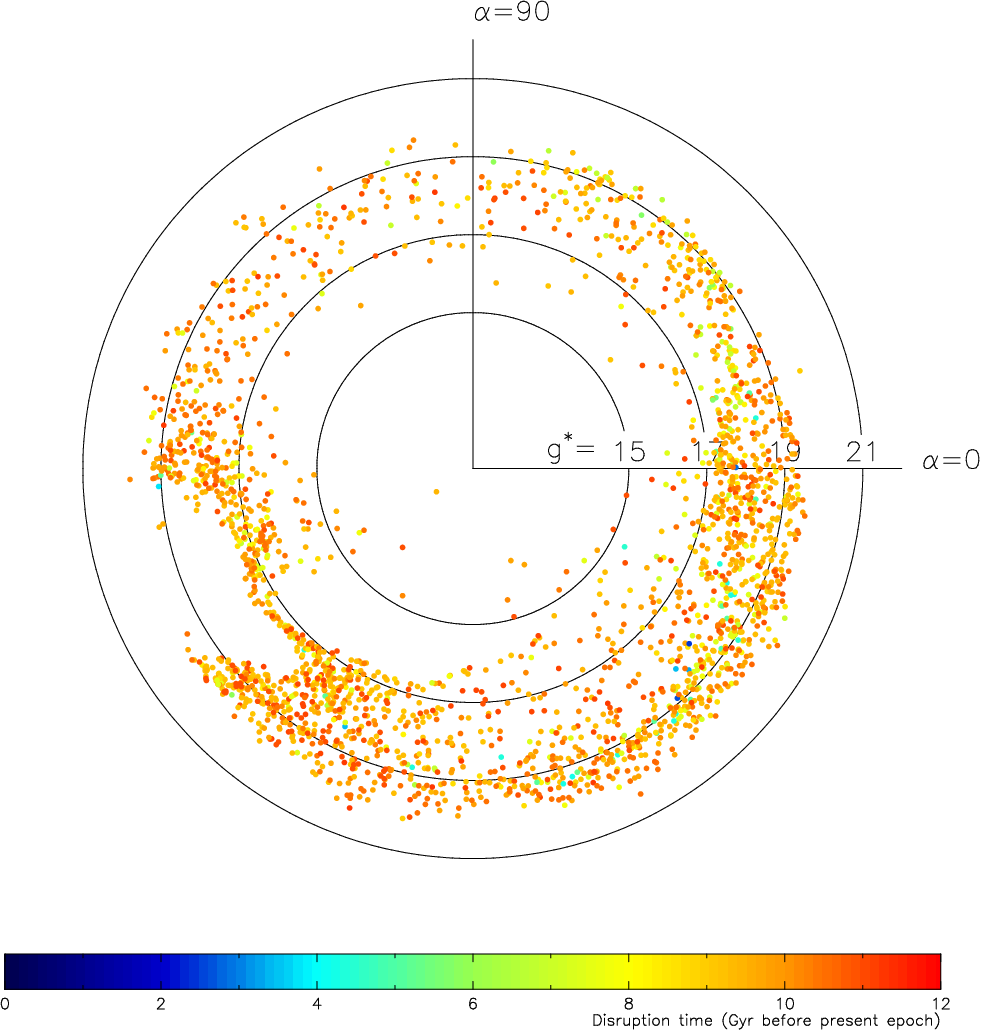}}
\figcaption{Identical to  Figure~2, except that  the data is derived  from a
simulation in  a $q_m=0.5$  flattened halo.  Note  the marked  difference to
Figure~2,  especially in  the right  ascension range  $270^\circ <  \alpha <
0^\circ$.   Further  SDSS  data  in   that  region  will  greatly  help  the
discrimination  of  competing  models  of  the dynamical  evolution  of  the
Sagittarius dwarf galaxy,  and of the structure of the  outer Halo. The hole
in the particle distribution near $210^\circ$ is due to the structure of the
tube orbits populated by the particles (see Figure~10 of paper~1).}
\end{figure}


\begin{thebibliography}{}
%
\def\mnras{\sl Mon. Not. R. astr. Soc.}
\def\apj{\sl Astrophys. J.}
\def\aj{\sl Astr. J.}
%
\bibitem[Azzopardi \& Lequeux(1992)]{azzopardi}
	Azzopardi, M. \& Lequeux, J. 1992,
	in  Stellar  Populations  of  Galaxies, IAU Symp. 149, p201,  
        Eds  B.  Barbuy  \& A. Renzini, Kluwer
%
\bibitem[Da Costa(1998)]{dacosta}
	Da Costa, G. 1998, in Proc.  VIIIth Canary Islands Winter School, p351,
	Eds. A. Aparacio \& A. Herrero, CUP
%
\bibitem[Dehnen \& Binney(1998)]{dehnen}
	Dehnen, W. \& Binney, J. 1998, \mnras\ 294, 429
%
\bibitem[Ibata, Gilmore \& Irwin(1994)]{ibata94}
	Ibata, R., Gilmore, G. \& Irwin, M. 1994, Nature 370, 194
%
\bibitem[Ibata et al. (1997)]{ibata97}
	Ibata, R., Wyse, R., Gilmore, G., Irwin, M. \& Suntzeff, N. 1997,
	\aj\ 113, 634
%
\bibitem[Ibata et al.(2000)]{ibata00}
	Ibata, R., Lewis, G., Irwin, M., Totten, E., Quinn, T.
        2000, astro-ph/0004011
%
\bibitem[Ivezi\'c et al.(2000)]{ivezic}
	Ivezi\'c, Z., \etal\ 2000, astro-ph/0004130
%
\bibitem[Johnston et al.(1999)]{johnston}
	Johnston, K., Majewski, S., Siegel, M., Reid, I., Kunkel, W.
	1999, \aj\ 118, 1719
%
\bibitem[Layden \& Sarajedini(2000)]{layden}
	Layden, A. \& Sarajedini, A. 2000, \aj\ 119,1760
%
\bibitem[Moore et. al(1999)]{moore}
	Moore, B, Ghigna, S., Governato, F., Lake, G., Quinn, T., Stadel, J., Tozzi, P.
	1999, \apj\ 524, 19L
%
\bibitem[Morrison et al.(2000)]{morrison00}
	Morrison, H., Mateo, M., Olszewski, E., Harding, P., Dohm-Palmer,
	R., Freeman, K., Norris, J., Morita, M. 2000, astro-ph/0001492
%
\bibitem[Totten \& Irwin(1998)]{totten98}
	Totten, E. J. \& Irwin, M. J. 1998, \mnras\ 294, 1
%
\bibitem[Totten, Irwin \& Whitelock(2000)]{totten00}
	Totten, E. J., Irwin. M. J. \& Whitelock P. 2000, \mnras (in press) 
        (astroph-0001113)
%
\bibitem[Whitelock, Irwin \& Catchpole(1996)]{whitelock96}
        Whitelock, P., Irwin, M., Catchpole, R. 1996, New Astron. 1, 57
%
\bibitem[Yanny et al.(2000)]{yanny}
	Yanny, B., \etal\ 2000, astro-ph/0004128
%
\end{thebibliography}
\end{document}